\documentclass[twocolumn,a4paper,floatfix]{revtex4}

\usepackage{graphicx}
\usepackage{multirow}
\usepackage{amsmath}
\usepackage{color}
\usepackage{epsfig}
\usepackage{epstopdf}

\begin{document} 
\title{Phase diagram of the ST2 model of water}

\author{Frank Smallenburg$^{a}$$^{\ast}$,
\thanks{$^\ast$Corresponding author. Email: f.smallenburg@gmail.com
\vspace{6pt}} 
Peter H. Poole$^b$, and Francesco Sciortino$^c$\\
\vspace{6pt}  
$^{a}${\em{Institut f\"ur Theoretische Physik II: Weiche Materie, Heinrich-Heine Universit\"at D\"usseldorf, Universit\"atstrasse 1, 40225 D\"usseldorf, Germany}};\\
$^{b}${\em{Department of Physics, St. Francis Xavier University, Antigonish, Nova Scotia B2G 2W5, Canada}};\\
$^{c}${\em{Dipartimento di Fisica, Universit\`a di Roma {\em La Sapienza}, Piazzale A. Moro 5, 00185 Roma, Italy}
\vspace{6pt} }
}

\begin{abstract}
We evaluate the free energy of the fluid and crystal phases  for the ST2 potential [F.H. Stillinger and A. Rahman, J. Chem. Phys. {\bf 60}, 1545 (1974)] with reaction field corrections for the long-range interactions.  We estimate the phase coexistence boundaries in the
temperature-pressure plane, as well as the gas-liquid critical point and gas-liquid coexistence conditions.  
Our study frames the location of the previously identified liquid-liquid critical point relative to the crystalline phase boundaries, and opens the way for exploring crystal nucleation in a model where the metastable liquid-liquid critical point is computationally accessible.
\end{abstract}

\maketitle

\section{Introduction}

The thermodynamic behavior of water at low temperatures is unconventional. Several quantities, e.g. the isobaric density $\rho$, the isothermal compressibility $K_T$, and the constant-pressure specific heat $C_P$, are characterized by
non-monotonic temperature or pressure dependence~\cite{phystoday2003}.
Over the past decades, the anomalous behavior of these quantities has attracted the attention of numerous researchers. In 1992, a numerical investigation of the equation of state (EOS) suggested the  presence of  a liquid-liquid (LL) critical point~\cite{poole1992phase} in  the 
ST2 model~\cite{ST2}, an interaction potential that describes water as a classical, rigid, non-polarizable molecule.   
The presence of a LL critical point, located in the supercooled region, provides an elegant explanation of the thermodynamic anomalies that characterize liquid water and which become more pronounced close to such a critical point~\cite{widomlinePNAS}.
 
The conceptual novelty of a one-component system with more than one liquid phase
has stimulated the scientific community to deeply probe the physical
origin of this phenomenon~\cite{mishima1998relationship, soper2000structures, katayama2000first, kurita2004critical, taschin2013evidence, pallares2014anomalies, amann2013water, azouzi2013coherent, sellberg2014ultrafast}.
It is now clear that a LLCP, while common in  tetrahedral network-forming liquids~\cite{saikapre2001,sastrynature,starrinterpenetrating,abascal2010widom,franknphys2,starrsoftmatter2014}, can also be observed in complex one-component fluids when the (spherically symmetric) interaction potential   generates two competing length scales~\cite{jagla1999core,hes26,hes50,galloprl}.
In the last few years the interest has shifted towards the interplay between
the liquid-liquid critical point and crystal nucleation~\cite{limmerchandler2011, palmer2014metastable, franknphys2,bagchi2014,buhariwalla2015}. Indeed, in experiments, 
crystallization has so far prevented direct observation of this phenomenon in a one-component bulk system.
Only recently have computer simulations demonstrated the possibility of generating a
thermodynamically stable liquid-liquid critical point (as opposed to a metastable one) in models of network-forming liquids~\cite{franknphys2,starrsoftmatter2014}.

Accurate information on the phase coexistence boundaries between disordered and 
ordered phases is relevant not only to establish the thermodynamic fields of stability
of the different phases, but also as a reference for estimating when the liquid
becomes metastable. In turn, this has relevance for estimating when the barrier
to crystallization becomes finite and how rapidly the barrier decreases on
supercooling~\cite{flaviojcp}.  
Except for one early report focussing on the liquid-ice~I$_h$ boundary~\cite{rehtanz}, 
none of the coexistence lines between the gas, liquid, and the many phases of crystalline ice have been accurately determined for the ST2 model.
In this article we fill this gap and evaluate these coexistence boundaries by calculating the
fluid chemical potential (via thermodynamic integration) and the crystal
chemical potential (via the Frenkel-Ladd method~\cite{frenkladd}, extended to molecules~\cite{Vegajpcm}). We test
several crystals (ice I$_h$, I$_c$, VI, VII, and VIII)
and find that in the region of pressure where  thermodynamic 
anomalies appear (e.g. near the lines of maxima of $C_P$ and $K_T$) ice I$_h$ and I$_c$ have the same free energy within our numerical precision.  Unexpectedly, we discover that for the ST2 model,  on increasing pressure, the stable phase is a dense tetragonal crystal
with partial proton order. This structure has a free energy about 0.4 $k_BT$  
lower than
ice VII,  the structure obtained by interspersing two I$_c$ lattices. (Here $T$ is the temperature and $k_B$ is the Boltzmann constant.)
We also evaluate the (metastable) line of coexistence for the recently reported
ice $0$ lattice~\cite{ice0,ice0qm}, a structure which could act (according to the
Ostwald rule) as the intermediate phase in the process of nucleating the stable
ice I$_{h/c}$ crystal from the fluid. For completeness, we  determine the
location of the gas-liquid critical point, which is found to be at $T_c= 558.0 \pm 0.3 $K and $\rho_c=0.265 \pm 0.005$ g/cm$^3$.
   
\section{Model and simulation methods}\label{sec:methods}

We study, via Monte Carlo (MC) simulations,  the original ST2 potential as defined by Rahman and Stillinger~\cite{ST2}, with reaction field corrections to approximate the long-range contributions to the electrostatic interactions.  ST2 models water as a rigid body with an oxygen atom at the center and four charges $q=\pm 0.4 e$ (where $e$ is the electron charge), two positive and two negative, in a tetrahedral geometry.  The distances from the oxygen to the positive and negative charges are 0.1 and 0.08~nm respectively.  The oxygen-oxygen interaction is modeled via a standard Lennard-Jones potential truncated at $2.5 \sigma_{LJ}$, with $\sigma_{LJ}=0.31$~nm and $\epsilon_{LJ}=0.316 94$ kJ/mol.  The Lennard-Jones 
residual interactions are handled through standard long-range corrections, i.e. by assuming that
the radial distribution function is unity beyond the cutoff.  The charge-charge interactions are smoothly switched off both at small and large distances via a tapering function, as in the original model~\cite{ST2}.  Complete details of the simulation procedure are as described in Ref.~\cite{poole1992phase}. In the following, we use $\sigma=1$ nm as unit of length. 

\subsection{Thermodynamic integration: Fluid free energy}

To evaluate the fluid free energy we perform thermodynamic integration
along a path of constant reference density $\rho_\mathrm{ref}$ for a modified pair potential,
\begin{equation}
V = \min(V_{ST2},200 \mathrm{\; kJ/mol}).
\end{equation}  
This potential coincides with the ST2 potential for all intermolecular distances and orientations where 
$V_{ST2}<200$~kJ/mol, and is constant and equal to 200~kJ/mol otherwise. Note that in the temperature range where we investigate the phase behavior,
molecules never approach close enough to reach this limit.
In this way, the divergence of the potential energy for configurations in which some intermolecular separations vanish 
(which would otherwise be probed at very high
temperatures) is eliminated and the infinite temperature limit is properly 
approximated by an ideal gas of molecules at the same density.

\begin{figure} 
\begin{center}
\includegraphics[width=0.45\textwidth]{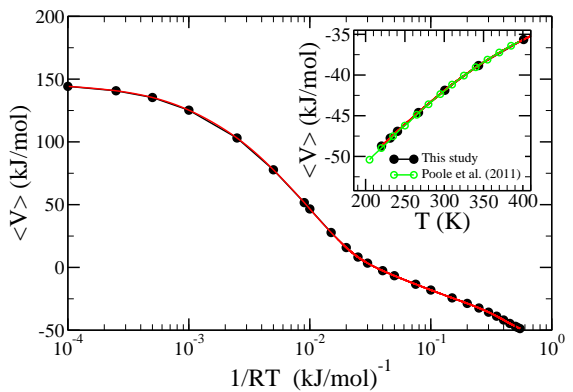}
\end{center}
\caption{Average pair potential energy $\left \langle V \right\rangle$ vs. $(RT)^{-1}$ at $\rho=1.0$ g/cm$^{3}$ (with $R$ the ideal gas constant). Symbols are MC data, and the line is the spline function used in the numerical integration.
The inset shows the same data as a function of $T$ and compares them to previously published data for the ST2 potential~\cite{poole2011dynamical}. }
\label{fig:tifluid}
\end{figure}

The fluid free energy (per particle) is calculated as
\begin{equation}
\beta f_{ST2}^\mathrm{fluid}(\beta,\rho_\mathrm{ref})=\beta f_\mathrm{ig}(\beta,\rho_\mathrm{ref}) + \int_0^{\beta}
\left\langle V(\beta,\rho_\mathrm{ref}) \right\rangle d\beta,
\end{equation}
where $\beta =1/k_B T$  and $\beta f_\mathrm{ig}(\beta, \rho) = \log(\rho_n \sigma^3) - 1$ is the ideal gas free energy and 
$\rho_n$ is the number density.  Fig.~\ref{fig:tifluid} shows the average modified pair potential energy $\left \langle V(\beta,\rho)\right\rangle$ and the
interpolating (spline) continuous curve used to numerically evaluate the integral. 
The free energy at different densities along a constant-$T$ path  is evaluated via thermodynamic integration of the equation of state
\begin{equation}
\beta f_{ST2}^\mathrm{fluid}(T,\rho_n)= \beta f_{ST2}^\mathrm{fluid}(T,\rho_{n,\mathrm{ref}}) + \int_{\rho_{n,\mathrm{ref}}}^{\rho_n}
\frac{\beta P(\rho_n')}{\rho_n'} d \ln(\rho_n'),
\end{equation}
where $P(\rho_n)$ is the equation of state for the pressure $P$ at fixed $T$.

\subsection{Crystal free energy}

To evaluate the free energy of a selected crystalline structure
we follow the methodology reviewed in Ref.~\cite{Vegajpcm}. 
We  define  an Einstein crystal in which each  molecule interacts, in addition to the ST2 potential, with a 
Hamiltonian, composed of a 
translational ($H_\mathrm{trans}$) and a rotational ($H_\mathrm{rot}$) part, that attaches each molecule to a reference position and orientation. 
For each particle we  define two unit vectors: the (normalized) HH vector and dipole vector, named respectively $\vec a$ and $\vec b$.   The reference configuration is defined by the reference position of the oxygen atom ${\bf r_0}$ and
the reference position of  $\vec a$ and $\vec b$~\cite{Vegajpcm,noya}. 
In the following we indicate with ${\bf r}-{\bf r}_0$ the 
  displacement of a particle located at ${\bf r}$  from
its  reference position, and with  $\phi_a$ and
$\phi_b$ the angles between  $\vec a$ and $\vec b$
and their reference values. More precisely, 
\begin{equation}
H_\mathrm{Einstein}= H_\mathrm{trans}+H_\mathrm{rot}
\end{equation}
with 
\begin{equation}
H_\mathrm{trans}= \lambda_t ({\bf r- r_0})^2/\sigma^2
\end{equation}
and
\begin{equation}
H_\mathrm{rot}= \lambda_r \left[   \sin^2 \phi_a + \left(\frac{\phi_b}{\pi}\right )^2 \right].
\end{equation}
Here $\lambda_t$ and $\lambda_r$ indicate the strength  of the coupling to the reference configuration.
Again following Ref.~\cite{Vegajpcm}, 
 the free energy (per particle) of a crystal structure $f^\mathrm{xt}$, in the limit of large $\lambda_r$ and $\lambda_t$ is calculated as,
\begin{equation}
\beta f^\mathrm{xt} =  \beta f_1 + \beta f_2 + \beta f_3 + \beta f_4 + \beta f_5 + \beta f_6
\end{equation}
where, indicating with $N$ the number of molecules in the system, 
\begin{eqnarray}
\beta f_1 &=& -\frac{1}{N} \ln  \left[ \left( \frac{ \pi}{\beta \lambda_t} \right)^{\frac{3(N-1)}{2}} N^\frac{3}{2}  \frac{1}{\rho_n \sigma^3 } \right]\\
\beta f_2 &=&  -\ln{ \frac{\sqrt{\pi}}{4}} + 1.5 \ln(\beta \lambda_r) \nonumber\\
\beta f_3 &=& \int_0^{\lambda_t}   \left\langle\beta  H_{trans}\right\rangle_{\lambda} d \ln \lambda \nonumber\\
\beta f_4 &=& \int_0^{\lambda_r}   \left\langle \beta H_{rot}\right\rangle_{\lambda} d \ln \lambda \nonumber\\
\beta f_5 &=& - \frac{\ln \left\langle e^{-\beta V_{ST2}}\right\rangle_{\lambda_r,\lambda_t} }{N}\nonumber\\
\beta f_6 &=& \nonumber
\left\{
\begin{array}{ll} 
\ln[1.5]  & \text{(full proton-disordered crystal)} \\
 0 &\text{(proton-ordered crystal).} \end{array} \right.
\end{eqnarray}
The symbols $\left\langle H_\mathrm{rot}\right\rangle_{\lambda}$ and $\left\langle H_\mathrm{trans}\right\rangle_{\lambda}$ indicate
the average values of $H_\mathrm{rot}$ and $H_\mathrm{trans}$ calculated from a
MC simulation of particles interacting via the ST2 potential complemented by
$H_\mathrm{Einstein}$.   The symbol  $\left\langle e^{-\beta V_{ST2}}\right\rangle_{\lambda_r,\lambda_t}$
indicates the average value of $e^{-\beta V_{ST2}}$ (where $V_{ST2}$ is the
system ST2 potential energy) in a simulation in which the particles interact with each other via the ST2 potential and with
the Einstein Hamiltonian with values $\lambda_r$ and $\lambda_t$.
In all simulations carried out to perform the integration, the center of mass of the system
is kept fixed~\cite{frenkelsmith}.  

Finally, $\beta f_6$ indicates the contribution of proton disorder, evaluated according to Pauling's
estimate~\cite{pauling}. More recent calculations have essentially confirmed 
Pauling's value~\cite{berg2007residual}.

Table~\ref{table:data} reports the values of $\beta f_j$ for a few representative cases.

\begin{table*}
\centering
{
\begin{tabular}{|c|c|c|c|c|c|c|c|c|c|c|c|}
\hline
~ & T (K) & $\rho$  (g/cm$^3$) &  N & $\lambda_r$ (kJ/mol) & $\lambda_t$ (kJ/mol) &  $\beta f $ & $\beta f_1$ & $\beta f_2$  & $\beta f_3+\beta f_4$  & $\beta f_5$ & $\beta f_6 $ \\
\hline
Ice I$_h$ & 270 & 0.8715 & 21952 & $3 \cdot 10^4$ & $3 \cdot 10^6$ & -8.829 & 19.440    &  15.064  & -19.640     &  -23.282  & -0.410 \\
\hline
Ice I$_c$ & 270 & 0.8715 & 21952 & $3 \cdot 10^4$ & $3 \cdot 10^6$ & -8.829 & 19.440   &  15.064 & -19.658  &  -23.264 & -0.410 \\
\hline
Ice VI & 250 & 1.27356 & 8100 &     $3 \cdot 10^4$ & $3 \cdot 10^6$ & -10.772 & 19.678   &  15.305 & -20.349  &  -25.996 & -0.410 \\
\hline
Ice VII & 270 & 1.5804 & 21296 & $3 \cdot 10^4$ & $3 \cdot 10^6$ & -8.230 & 19.440   &  15.064 & -19.318  &  -23.006 & -0.410 \\
\hline
Ice VII$^*$ & 270 & 1.6250 & 6912 & $3 \cdot 10^4$ & $3 \cdot 10^6$ & -8.59 & 19.440   &  15.064 & -19.112  &  -23.567 & -0.410 \\
\hline
Ice VIII & 270 & 1.55645 & 1152 & $3 \cdot 10^4$ & $3 \cdot 10^6$ & -6.852 & 19.4185   &  15.064 & -19.061  & -22.274 & 0 \\
\hline
Ice 0 & 250 & 0.8494 & 29160 &      $3 \cdot 10^4$ & $3 \cdot 10^6$ & -10.399 & 19.555   &  15.180 & -19.983  &  -24.741 & -0.410 \\
\hline
Fluid &       270 & 1.002  & $268$ &   --- & ---                & -8.4411 &  --- &  --- & --- & --- & ---\\
\hline
\end{tabular}}
\caption{Free energy $\beta f$ of the fluid and crystal phases at selected points. The value for the residual entropy in $\beta f_6$ was taken from Ref. \cite{berg2007residual}. The columns marked $\beta f_i$ indicate the various contributions from the crystal free energy calculation. For the fluid, we used thermodynamic integration from an ideal gas at constant density, as explained in the text. }
\label{table:data}
\end{table*}%

\subsection{Grand canonical simulation: Gas-liquid phase coexistence}

To evaluate the gas-liquid coexistence and the location of the gas-liquid critical point,
we perform grand-canonical MC simulations to evaluate
at fixed $T$, volume $v$, and chemical potential $\mu$, the
probability $p$ of observing $N$ particles in the simulated volume.  To overcome the
large free energy barriers separating the gas and liquid phases we 
implement the successive umbrella sampling (SUS) technique~\cite{sus}.
Since this method has been applied previously to ST2~\cite{pccp} to estimate the
liquid-liquid coexistence conditions, and has been documented in detail in these
works, we refer the interested reader to the original literature.

\subsection{Proton position in the crystal structures}

\begin{figure} 
\begin{center}
\includegraphics[width=0.45\textwidth]{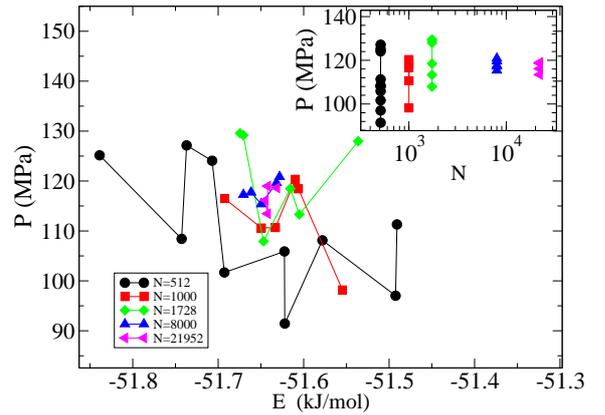}
\end{center}
\caption{Size effects associated with proton disorder in configurations 
with zero net dipole moment: Relation between 
the average pressure and the average energy of different proton-disordered configurations, for different number $N$ of 
molecules. For all cases,  the structure is ice I$_c$ at $T$=270 K  and $\rho=0.8715$ g/cm$^{3}$.
The inset shows the average pressure in different proton-disordered configurations
for different number of molecules. }
\label{fig:size}
\end{figure}

To generate proton-disordered crystals, such as ice I$_{h/c}$ and ice VII,
one needs to assign protons to the oxygens, located at the  lattice positions,
so as to satisfy the ice rules. To this end, we  first calculate a list of all bonded oxygen neighbours (where four bonds connect to each oxygen atom)  and then decorate the oxygen lattice by assigning the proton for each bond to one of the two bonded atoms,  iterating the following procedure:
 (i) Randomly select one oxygen with less than two  hydrogens and one of the remaining
    undecorated bonds emanating from the selected oxygen. 
 (ii) Randomly follow the path of undecorated bonds until the path loops back to the original oxygen.
 (iii) Decorate all bonds of the selected path with one proton each, associating the protons
  to the oxygens encountered in the path.   The procedure is iterated until all oxygens have two protons associated with them.  Paths in which the initial and final oxygen atoms coincide only via  periodic images produce a non-zero dipole moment and should be rejected if the net dipole moment of the cell is to vanish. 
  
 To account for all possible proton realizations one needs to investigate large systems
 or average over several configurations. Indeed, we find that 
 there is a significant correlation between the proton realization and the average potential 
 energy $E$ and average pressure $P$ (at constant volume). Fig.~\ref{fig:size} 
  correlates $P$ and $E$ for each realization, while the inset 
  shows 
  $P$ in different realizations for system sizes from
 $N=512$ to $N=21952$ molecules.  
  Only for 8000 or more particles is the variance between different realizations
 within a few MPa and a tenth of a kJ/mol, the tolerance required to allow for a precise determination
 of the thermodynamic variables entering into the free-energy calculation. 
 Unless otherwise stated, we have analyzed configurations with 8000 or more particles
 for all proton-disordered crystals.

\section{Results}
\subsection{Gas-liquid coexistence}

\begin{figure*} 
\begin{tabular}{lll}
a)& \hspace{0.05\textwidth} &b)\\
\includegraphics[width=0.45\textwidth]{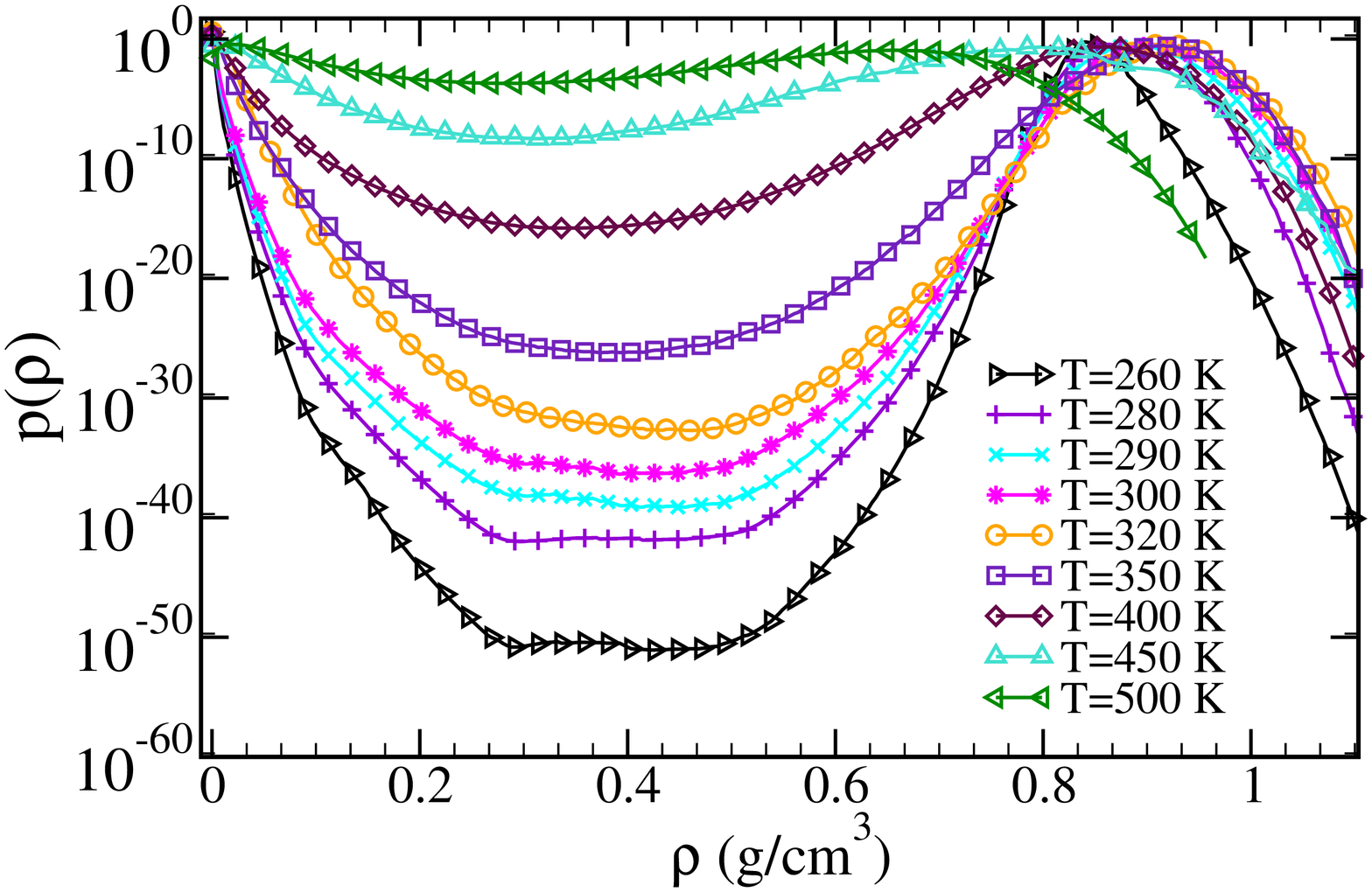} & & \includegraphics[width=0.45\textwidth]{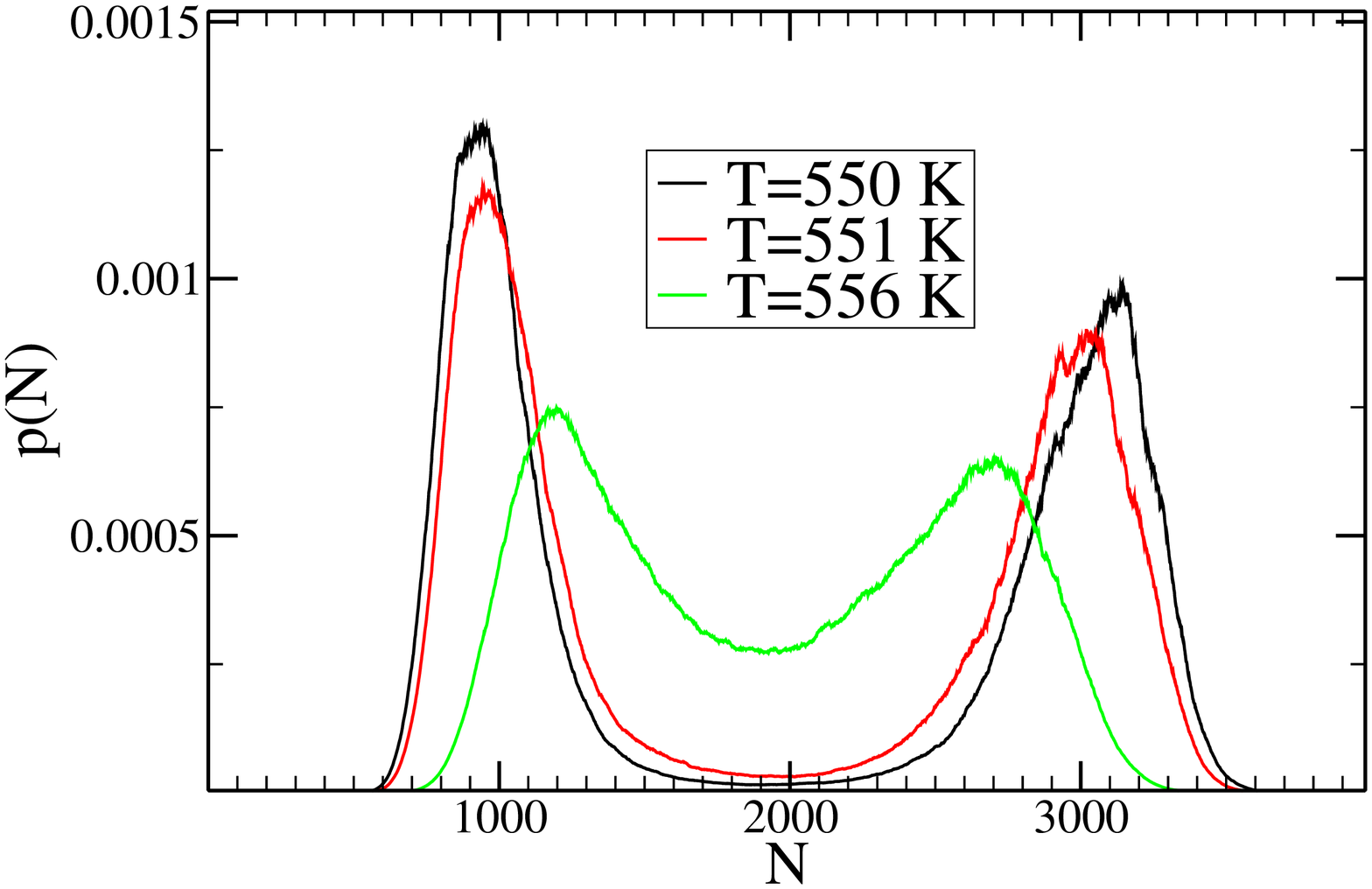} \\
c)& &d)\\
\includegraphics[width=0.45\textwidth]{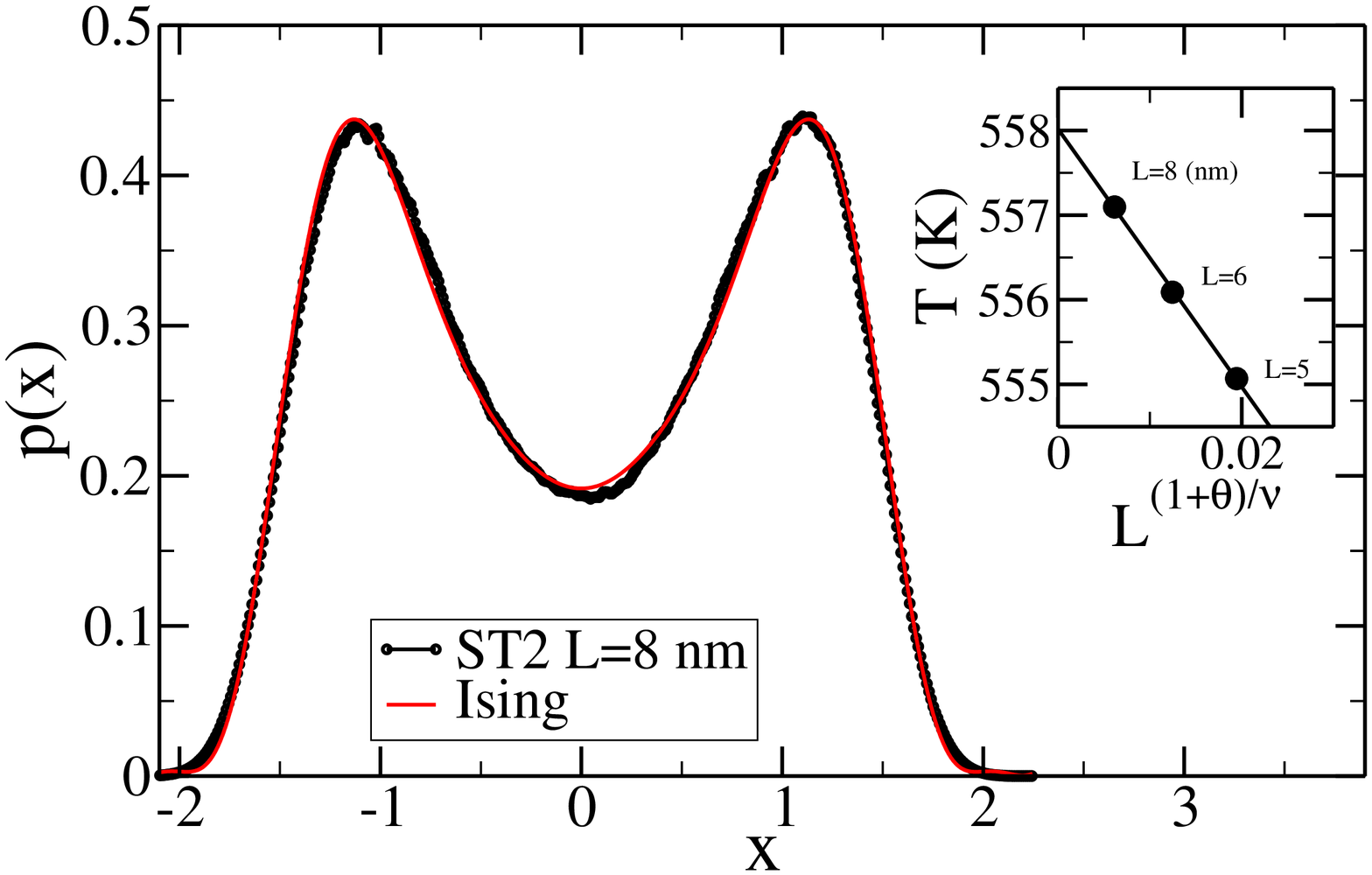} & & \includegraphics[width=0.45\textwidth]{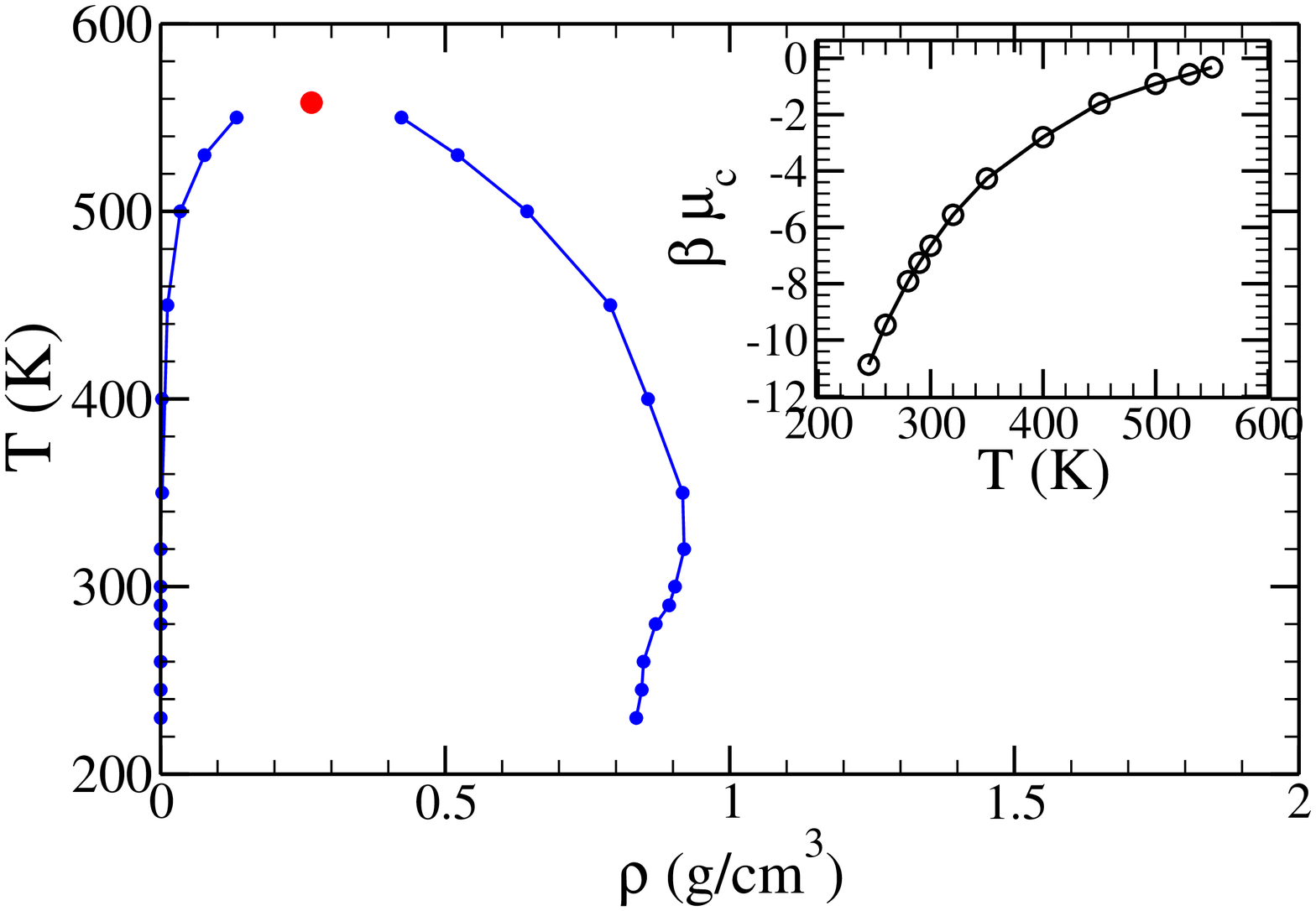}
\end{tabular}
\caption{Gas-liquid coexistence for the ST2 model. (a)  Distribution of the density fluctuations 
in gas-liquid coexisting states for different $T$ for system size $L=2$ nm.  (b) Same data as in (a) but for
$T$ close to the critical temperature $T_c$ and for $L=6$ nm.  (c) Comparison of the fluctuation in the
order parameter $x$ (a linear combination of $N$ and $E$) with the theoretical expression for the
three-dimension Ising model.  The inset shows the finite size scaling of the critical temperature.  (d) Resulting gas-liquid phase diagram in the  $T-\rho$ plane. The inset shows the values of the chemical potential along the coexistence. }
\label{fig:phase}
\end{figure*}
  
Fig.~\ref{fig:phase} shows the results of the SUS calculations. Panel (a) shows the
probability $p$ of finding $N$ particles at fixed $T$ and $v$ at 
the coexistence chemical potential $\mu_c$ for different $T$.   $\mu_c$  is 
evaluated by reweighting the histogram $p(N)$ with respect to $N$, such that the area below the gas and the
liquid peak is identical (0.5).   At low $T$, the probability minimum separating the two phases is more than 50 orders of magnitude lower than the peak heights, 
highlighting the need for a numerical technique (like SUS) that allows the 
observation of rare states.   Close to the critical point [panel (b)], the probability of exploring
intermediate densities between the gas and the liquid becomes significant and $p(N)$ [or $p(\rho)$]
assumes the characteristic shape typical of all systems belonging to the same universality class.
Panel (c) compares $p(N+sE)$, where $E$ is the potential energy of the configuration and $s$ is the so-called
mixing field parameter~\cite{wilding1995critical}, with the theoretical expression for the magnetization in the Ising model.
To reinforce the identification of the critical point with the Ising universality class, the inset shows
the finite size scaling of the critical $T$ (defined as the $T$, for each size, at which the fluctuations in
$N+sE$ are best fitted with the Ising form) as a function of $L^{(1+\theta)/\nu}= L^{-2.448}$, with $\theta =0.54$ and $\nu=0.630$~\cite{ferrenberg1991critical,pelissetto2002critical}.
The extrapolation to $L \rightarrow \infty$ suggests that the gas-liquid critical point for the
reaction field ST2 model is $T_c= 558.0 \pm 0.3 $K and $\rho_c=0.265 \pm 0.005$ g/cm$^3$.
Finally, panel (d) shows the gas-liquid coexistence in the $\rho-T$ plane.  A clear nose appears around $T=300$ K, 
signaling the onset of the  network of  hydrogen bonds (HB).  
Indeed,  strong directional  interactions  (such as the HB), impose a strong coupling between density and energy.
The formation of a fully bonded tetrahedral network (the expected thermodynamically stable state at low $T$)
requires a well-defined minimum local density, which for the present model is approximately $\rho=0.8$ g/cm$^3$. 
Hence, at low $T$, the density of the  network coexisting with the gas must approach this value.
For completeness, the inset in panel (d) reports the value of $\beta \mu_c$ along the coexistence line.

\subsection{Fluid-crystal coexistence}

\begin{figure} 
\begin{center}
\includegraphics[width = 0.45\textwidth]{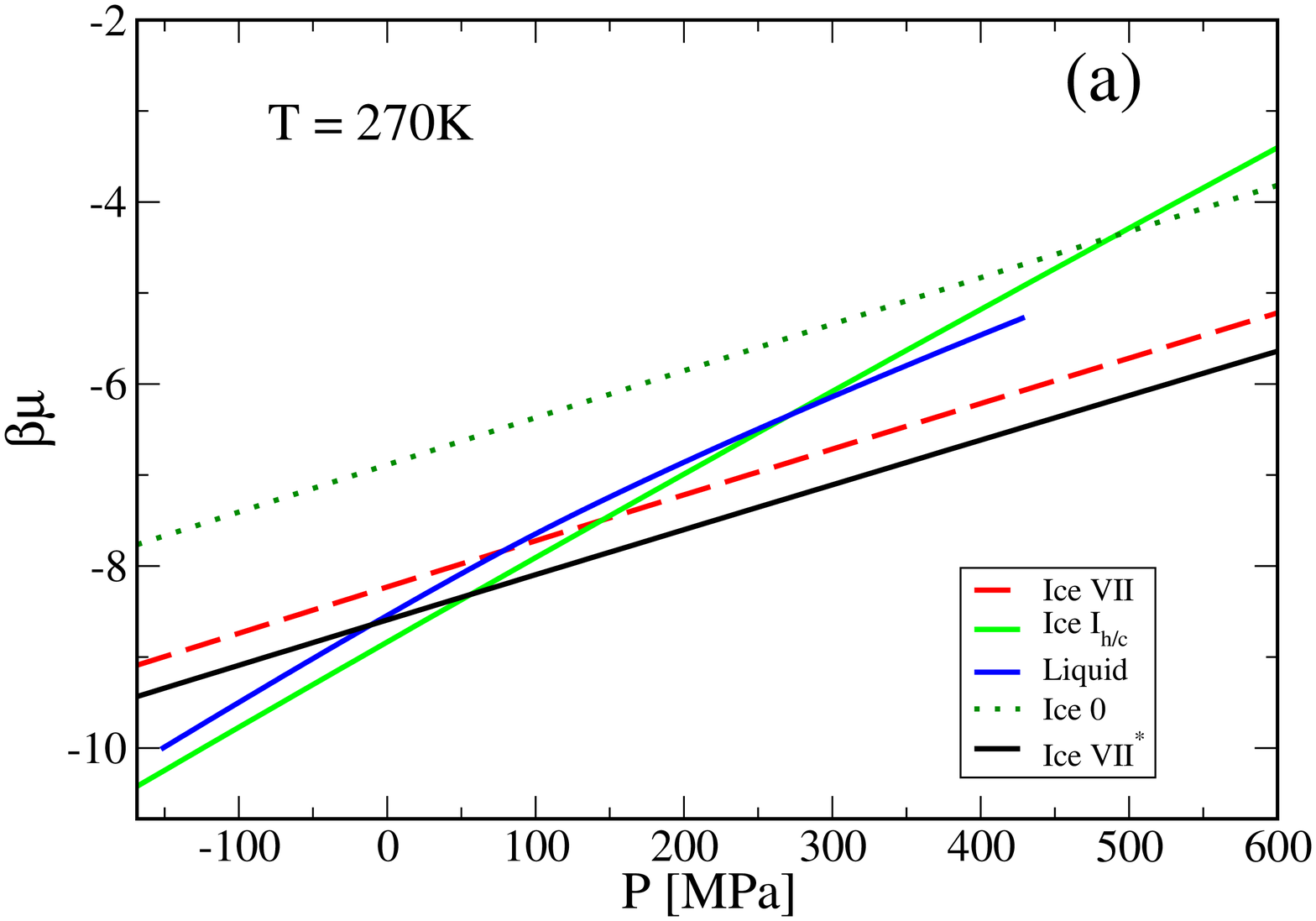}
\includegraphics[width = 0.45\textwidth]{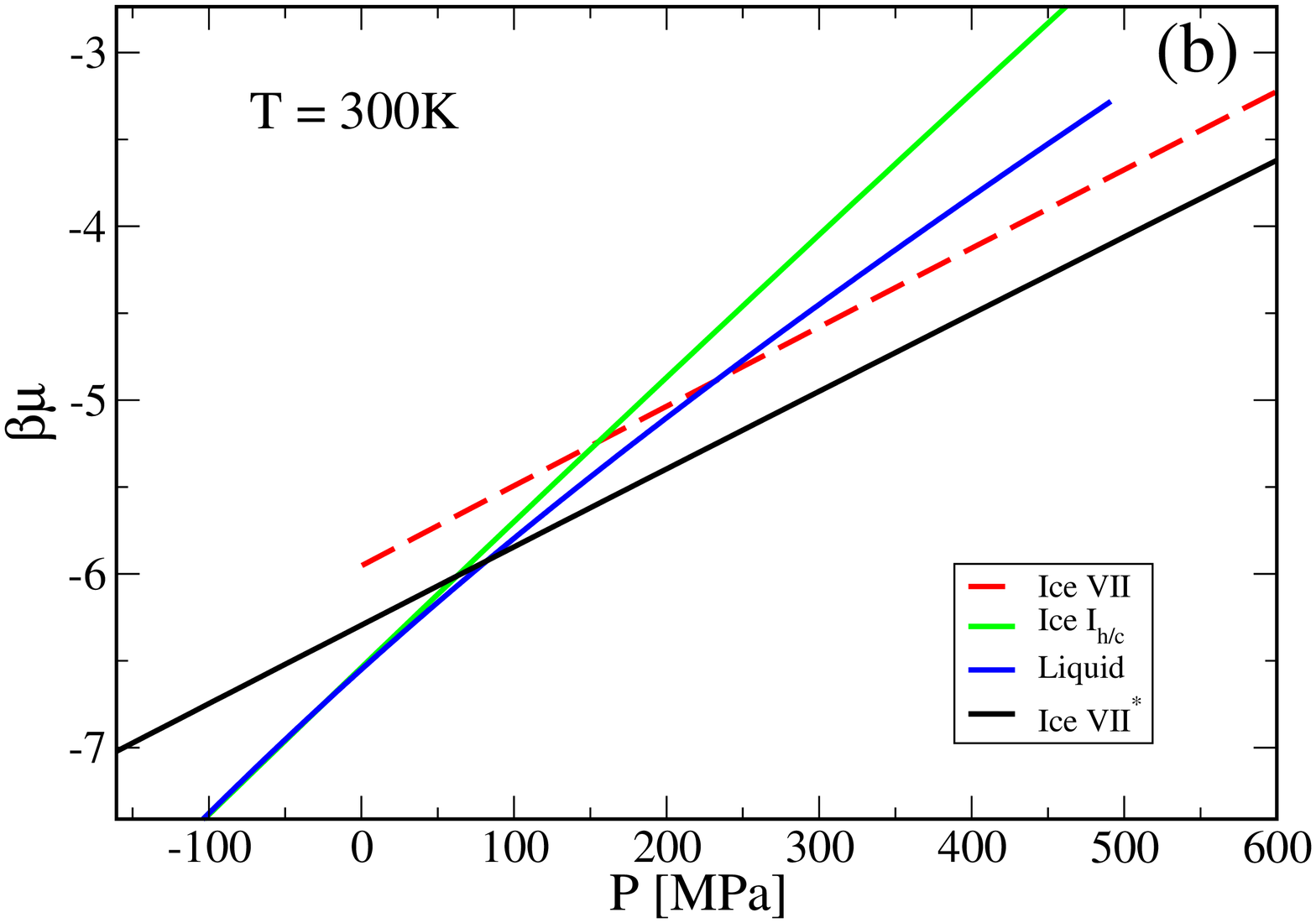}
\end{center}
\caption{Reduced chemical potential $\beta \mu$ as a function of pressure $P$  for competing phases at $T$=270~K and $T$=300~K. At each pressure, the phase with the lowest chemical potential is the stable one. Crossings indicate (metastable or stable) phase transitions.}
\label{fig:bmumany}
\end{figure}

We have investigated the stability of crystal phases that may coexist with the fluid at low $T$. In particular, we have determined the free energies of ices I$_c$, I$_h$, VI, VII, and VIII, as well as the recently proposed metastable ice 0 structure \cite{ice0}. Note that with the exception of ice VIII, all these phases have disordered hydrogen bonding.  Examples of our thermodynamic integration results are reported in 
Fig.~\ref{fig:bmumany}, where we plot   the reduced chemical potential  $\beta \mu \equiv \beta f + \beta P/\rho_n$ 
of different phases at two selected $T$. For each pressure interval, the lowest chemical potential phase is the 
thermodynamically stable one.   Intersections of different curves locate coexistence
points, either stable or metastable.    We then interpolate  the fluid and crystal free energies based on the equation of state
to  draw the coexistence lines in the phase diagram.

\begin{figure} 
\begin{center}
\includegraphics[width = 0.45\textwidth]{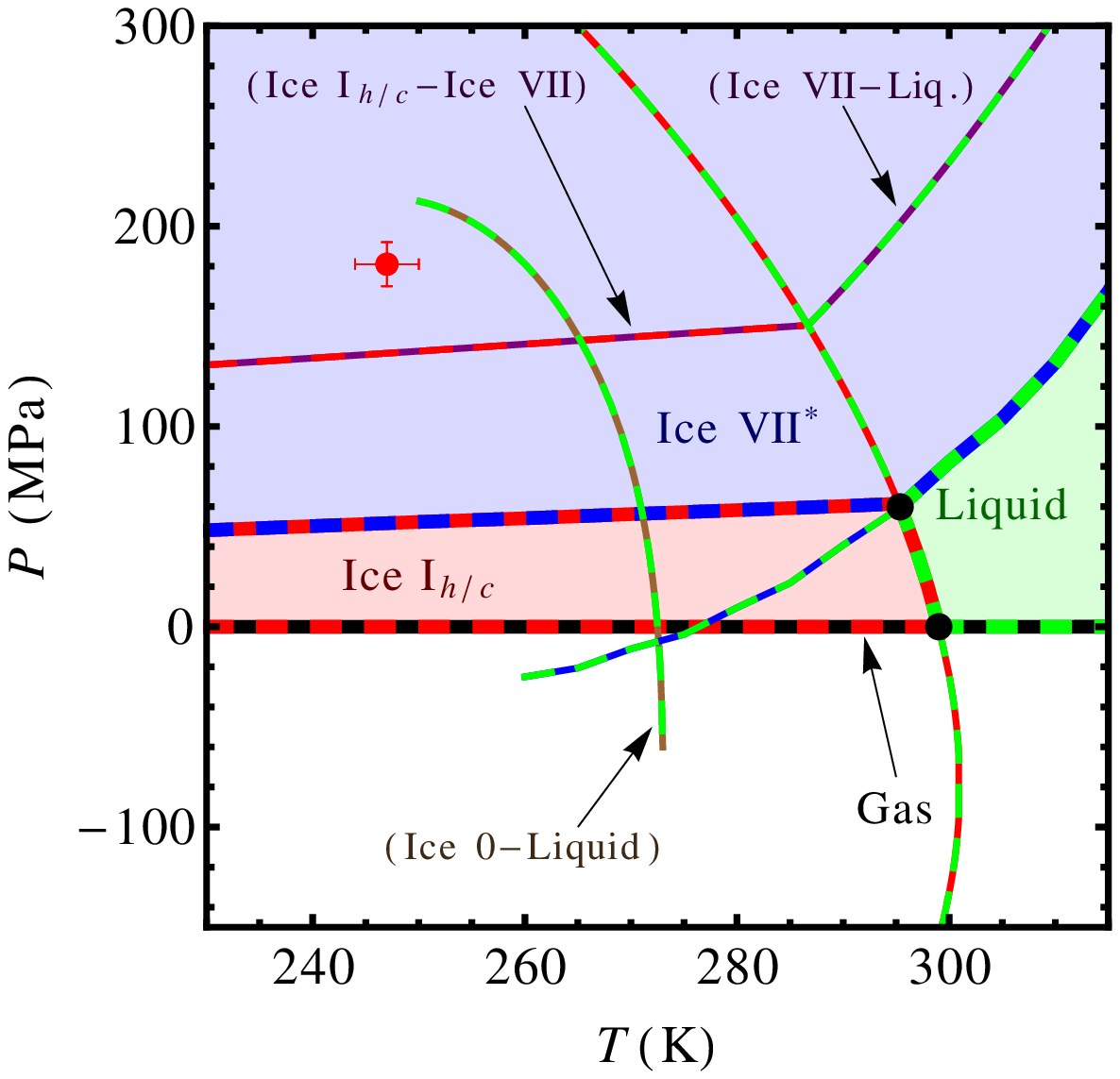}
\includegraphics[width = 0.45\textwidth]{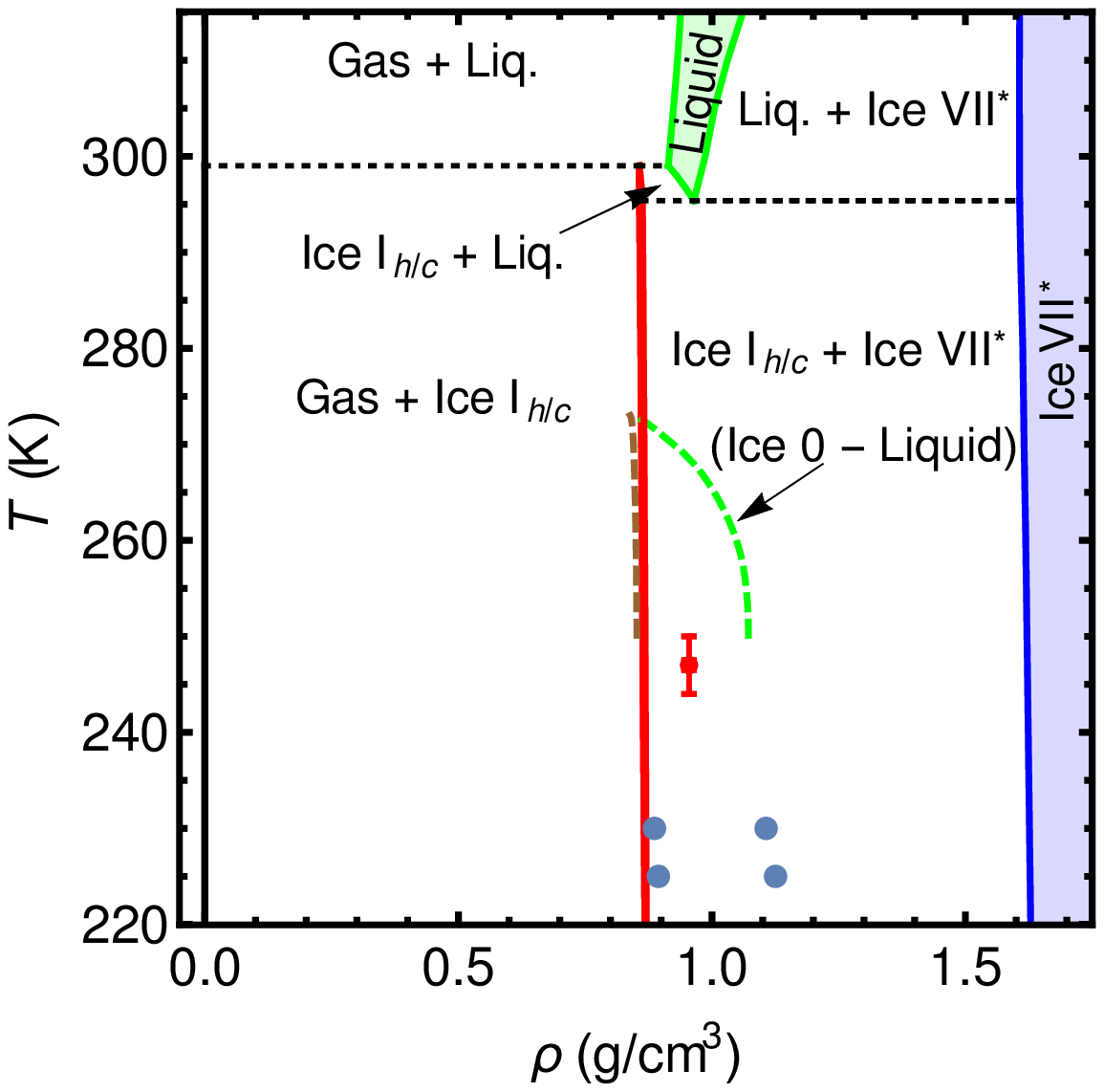}
\end{center}
\caption{(a)  Pressure-temperature phase diagram for the ST2 model with reaction field.  In this $P$-$T$ region, the stable phases are the  ice VII$^*$,  cubic or hexagonal  ice  (ice I$_{h/c}$), gas, and liquid.
The thick lines indicate phase boundaries between stable phases, while the thinner lines denote metastable phase transitions. 
The full circle indicates the location of the liquid-liquid critical point and its error bars, as estimated in Ref.~\cite{cpst2}. 
(b) Temperature-density representation of the same phase diagram.  White regions indicate two-phase coexistence. Filled regions indicate
areas of one-phase stability for the different phases.  
Solid lines indicate coexistence densities. 
Note that the one-phase stability field of ice I$_{h/c}$ is centred on the optimal crystal density and is very narrow, comparable in width to the coexistence lines.  
Horizontal dotted lines correspond to triple points.  We use the same colour-coding  as in (a).
The blue dots indicate the coexisting densities of the low and high density liquids, from Ref.~\cite{pccp}.  The red square is the estimated location of the LL critical point.}
\label{fig:ptphase}
\end{figure}

The complete phase diagram is reported in Fig.~\ref{fig:ptphase}.  At low $T$ and low $P$, the most stable crystal structure is the ice I lattice. From our simulations, the cubic (I$_c$) and hexagonal (I$_h$) ice structures have the same free energy within our numerical accuracy. At positive pressures, the liquid phase coexisting with ice I is always denser than  ice, and as a result, the melting temperature of  ice I  decreases with increasing $P$. At negative $P$ (near $P = -80$ MPa), the ice I and liquid phases coexist at the same density, and the melting temperature reaches a maximum. We note that we have confirmed the ice I$_{h/c}$ melting temperature calculated via thermodynamic integration 
at two separate pressures using direct coexistence simulations, and find good agreement. We note  that for the ST2-Ewald model,\
the  melting temperature of I$_c$ at the single pressure of 260 MPa was estimated to be around 274 K, consistent with the present estimate~\cite{palmer2014metastable}.

\begin{figure}
\begin{center}
\includegraphics[width = 0.45 \textwidth]{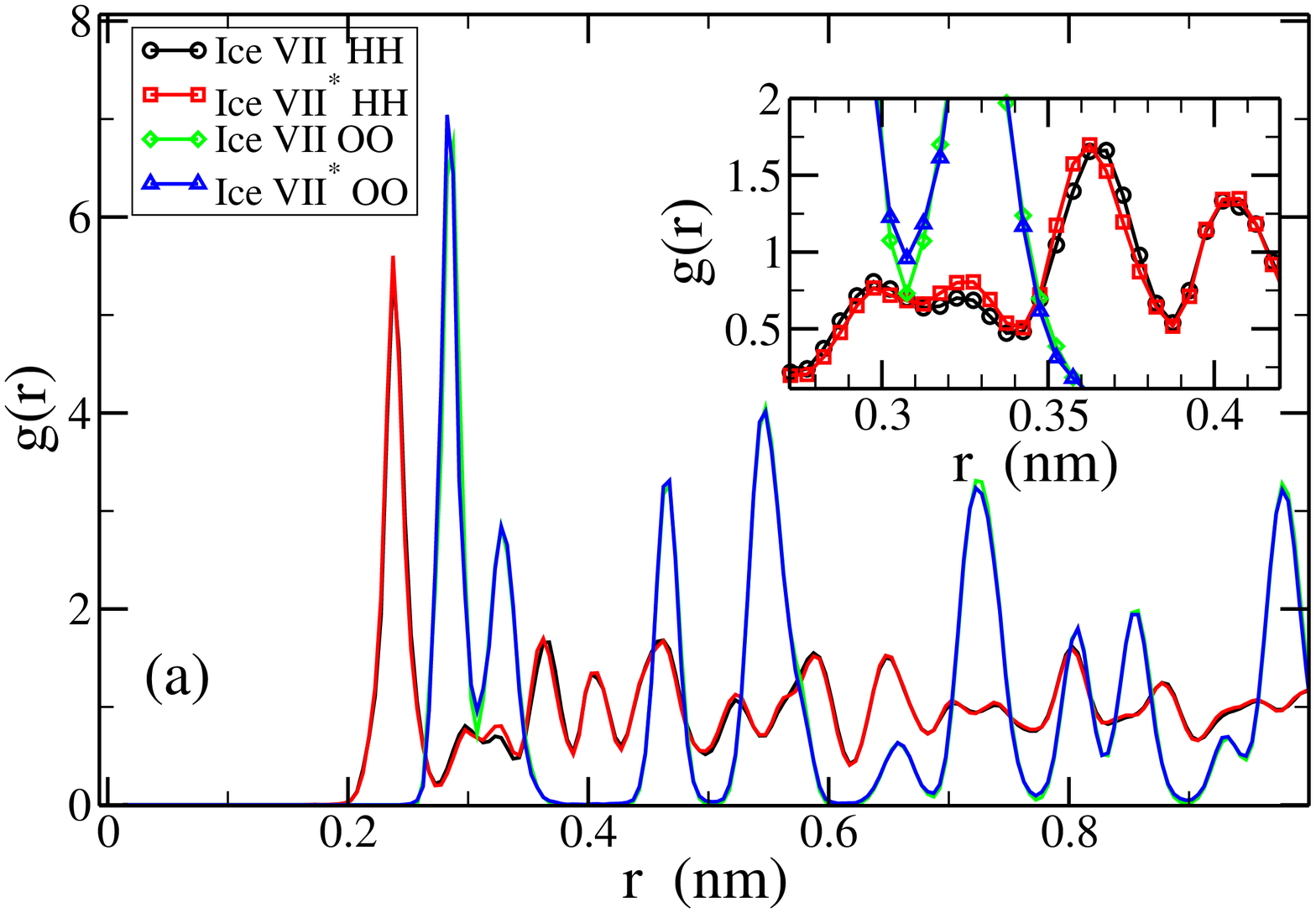}\\
\includegraphics[width = 0.45 \textwidth]{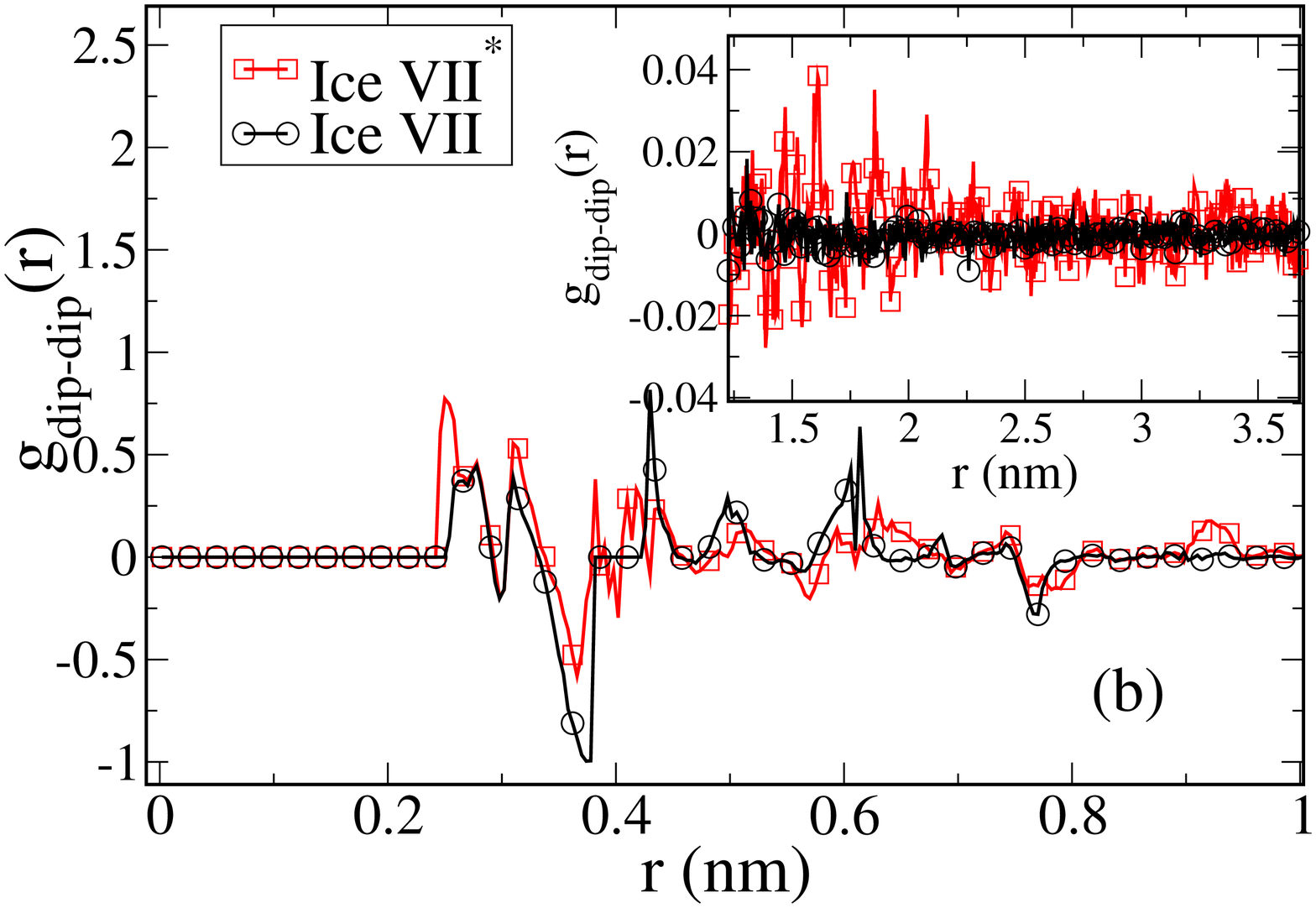}
\end{center}
\caption{(a) Radial distribution function and (b) dipole dipole correlation function for the fully proton-disordered ice VII structure and the regrown ice VII$^*$ crystal, at $\rho=1.673$ g/cm$^3$, in the inherent structure. Note the slight extra correlation in $g_\mathrm{HH}(r)$ and $g_\mathrm{dip-dip}(r)$ in the region 2~\AA $< r <$ 4~\AA.}
 \label{fig:BCCcorrelation}
\end{figure}

At high pressure, the main candidate structures are the proton-ordered ice VIII structure, and the proton-disordered ice VII structure. Both structures consist of two interpenetrating I$_c$ lattices (somewhat distorted in the case of proton-ordering), where the oxygen positions form  a BCC lattice. According to our free-energy calculations, the disordered ice VII is the more stable one in the region where coexistence with the fluid might occur. However, when trying to confirm the accuracy of our predicted liquid-ice VII coexistences using direct coexistence simulations, we observed crystal growth at temperatures significantly above the   melting temperature predicted from free energy calculations. The newly grown parts of the crystal still display the BCC topology of the oxygen atoms, but  the crystal  shrinks by a few percent in the direction perpendicular to the growth direction, leading to a slight distortion of the lattice, that we refer to in the following as ice VII$^*$.
 As this distortion does not occur in fully disordered ice VII, we attribute the unexpectedly high stability of the  ice VII$^*$ lattice to the emergence of partial proton ordering, which  decreases  the crystal free energy. 
To confirm this, we created a fully regrown ice VII$^*$  configuration by alternately melting and regrowing the two halves of an ice VII configuration in an elongated simulation box. When measuring the proton-proton and dipole-dipole correlation functions for both the original ice VII structure and the regrown ice VII$^*$, we see only minor changes in the proton-proton correlation function in the region 3~\AA $< r <$ 4~\AA~[see 
Fig.~\ref{fig:BCCcorrelation}(a)].  In contrast, the  dipole-dipole correlation function~[see 
Fig.~\ref{fig:BCCcorrelation}(b)]
shows   significant additional signal   which although weak, extends up to long spatial scales. Using the Frenkel-Ladd method, we calculate the free energy of this configuration (assuming full proton disorder), and find that it is indeed lower than that of the original crystal by $\approx 0.4~k_B T$ per particle, confirming that the lower melting temperature observed in our direct coexistence simulations can be attributed to the (slight) change in crystal structure.  The difference in free energy mainly results from the lower potential energy of the regrown crystal. We note here that partial proton ordering would reduce the contribution of the residual entropy to the free energy of the crystal, causing us to underestimate the ice VII$^*$ free energy. On the other hand, the presence of defects in the system is expected to cause an overestimate in the crystal free energy. It is thus not {\it a priori} obvious that this free energy can be used to predict coexistences. Nonetheless, comparing the melting temperature predicted from the free energy and equations of state of the regrown crystal with the melting temperature taken from the direct coexistence simulations, we find good agreement ($T \approx 320K \pm 5K$ at $P = 250$ MPa). Calculating the rest of the coexistence lines for this crystal using thermodynamic integration, we observe that   ice VII$^*$ has a significantly larger stability region than the original ice VII (see Fig. \ref{fig:ptphase}). 

We note that neither ice VI nor ice 0 are ever the most thermodynamically stable phase in the investigated region. 
As it may be relevant in future nucleation studies, we include the metastable coexistence line of the liquid with  ice 0  in the phase diagram (Fig. \ref{fig:ptphase}).

\section{Conclusions}
Recently, the ST2 potential has been at the centre of renewed interest in connection to the debate on the origin of the
liquid-liquid critical point~\cite{phystoday2003, franzese2001generic, PhysRevLett.91.155701, fuentevilla2006scaled, anisimov,holten2014}.  
This model exhibits known deficiencies in accurately modelling water properties, e.g. it
overemphasizes the tetrahedrality of the liquid structure, thus shifting all water anomalies to higher temperatures.
Despite these deficiencies, the ST2 model plays a key role as a prototype system in many studies related to the presence of a liquid-liquid critical point.  
We report here fundamental properties of the ST2 model, by evaluating the location of the 
gas-liquid critical point and the gas-liquid coexistence  curve, as well as the coexistence lines between the liquid
and several crystal structures, allowing us to map out the phase diagram of the ST2 model in the low-temperature regime. 
We find a stable ice I phase  at low pressure and temperature, with both the hexagonal and cubic stackings approximately equal in free energy. 
Differently from real water, the  high-pressure phase behavior of the model is dominated by a new crystal  whose growth is templated  by
the ice VII interface.  This ice VII$^*$  tetragonal crystal is composed of a lattice in which the oxygens have the same topology as  ice VII 
but in which the protons are not completely randomly distributed.  We have not been able to identify a small unit cell for this new crystal, but inspection of the HH
radial distribution function indicates minute  but observable differences in the region around $3.3$~\AA, accompanied 
by weak but long ranged correlations in the  dipole-dipole correlation function.
This structure, despite the small partial proton order,  has a significant lower potential energy than VII (approximately 1.2 kJ/mol). As a result, ice VII$^*$ 
is significantly more stable than the fully proton-disordered ice VII phase at all pressures and it
dominates the high-pressure phase behavior of the model.   The  liquid-liquid critical point for this model lies, according to the most recent estimates, inside the region of stability of the ice VII$^*$ crystal phase and is metastable with respect to ice I$_h$ or  I$_c$ as well as to ice VII.
Our results provide a starting point for the study of nucleation in the ST2 model, as well as for the exploration of modifications to the model~\cite{st2flex} that could make the liquid-liquid critical point more accessible~\cite{franknphys2}.

\section{Acknowledgments}
We thank L. Filion, A. Geiger, A. Rehtanz, and I. Saika-Voivod
for useful discussions.  We are honored to dedicate this study to 
Prof. Jean-Pierre Hansen, from whom we have learned  liquid state theory.

\bibliography{all.bib}

\end{document}